\documentclass[aps,nofootinbib,twocolumn,showkeys,floatfix]{revtex4}
\usepackage[utf8]{inputenc}
\usepackage[brazilian]{babel}
\usepackage[T1]{fontenc}
\usepackage{mathrsfs}
\usepackage{amsmath}
\usepackage{amssymb}
\usepackage{graphicx}
\usepackage{csquotes}
\usepackage{setspace}

\begin{document}

\title{Relatividade bem comportada: buracos negros regulares}
\author{J. C. S. Neves}
\email{nevesjcs@ime.unicamp.br}
\affiliation{Instituto de Matemática, Estatística e Computação Científica, Universidade
Estadual de Campinas \\ CEP. 13083-859, Campinas, SP, Brazil}

\begin{abstract}
\textbf{Resumo}: A recente observação das ondas gravitacionais corrobora uma das mais interessantes previsões da relatividade geral: os buracos negros. Pois as ondas gravitacionais detectadas pela colaboração LIGO ajustam-se muito bem dentro da teoria da relatividade geral como um fenômeno produzido pela colisão de dois buracos negros. Sendo assim, a realidade física dos buracos negros parece ainda mais inegável hoje. Embora, uma mais contundente prova sobre a existência de buracos negros seria dada pela observação do seu horizonte de eventos, aquilo que o define. Neste artigo, é indicado que somente o horizonte de eventos define um buraco negro. Em sua definição, não há menção à singularidade em seu interior. Mostrar-se-á, assim, que buracos negros sem singularidade são possíveis. Tais são hoje chamados de buracos negros regulares.   

\vspace{0.5cm}

\textbf{Abstract}: The recent observation of gravitational waves confirms one of the most interesting predictions in general relativity: the black holes. Because the gravitational waves detected by LIGO fit very well within general relativity as a phenomenon produced by two colliding black holes. Then the reality of black holes seems almost undoubted today. However, a stronger proof on the reality of black holes would be indicated by the observation of the event horizon, which is what defines it. In this article, it is indicated that only the event horizon defines a black hole. There is no mention to the singularity in its definition. Thus, it will be shown that black holes without a singularity are possible. Such black holes are called regular black holes.   
\end{abstract}


\maketitle

\section{Introdução}
A colaboração LIGO (\textit{Laser Interferometer Gravitational-Wave Observatory}) anunciou o mais impactante resultado em física no ano de 2016: a detecção das ondas gravitacionais \cite{LIGO}. Sendo uma previsão da relatividade geral,\footnote{O artigo original, escrito por Einstein em alemão, onde surge o conceito de ondas gravitacionais na teoria da relatividade geral, é a ref. \cite{Einstein}. Veja também \cite{Cattani-Bassalo}, onde comentários sobre a recente detecção das ondas gravitacionais são feitos na \textit{Revista Brasileira de Ensino de Física}.} feita logo após Albert Einstein publicar sua teoria, tivemos que aguardar cerca de um século para recebermos essa tão esperada confirmação. E tal confirmação abre as portas para uma provável nova área na ciência, a física das ondas gravitacionais. Se a astronomia, astrofísica e cosmologia dependeram da radiação eletromagnética para se desenvolverem até aqui, com as ondas gravitacionais um novo tipo de radiação --- a radiação gravitacional --- entra em cena, apresentando-nos o mundo a partir de um novo olhar ou perspectiva. 

Com os resultados da colaboração LIGO sobre as ondas gravitacionais, outro resultado no mesmo experimento, tão importante quanto, surge: a detecção de buracos negros. Conforme relatado pela colaboração, a detecção das primeiras ondas gravitacionais foi possível pois foram geradas pela colisão de dois buracos negros. E buracos negros são uma previsão da teoria da relatividade geral tão antiga quanto as ondas gravitacionais. O primeiro deles foi proposto ainda em 1916. Foi o físico alemão Karl Schwarzschild \cite{Schwarzschild} quem o propôs, num famoso artigo à academia prussiana de ciências, onde as equações do campo gravitacional (então recentemente propostas por Einstein na relatividade geral) de uma massa pontual no vácuo foram resolvidas. A solução hoje é conhecida como solução de Schwarzschild, em homenagem ao seu autor. Tal solução pôde ser interpretada (e foi depois) como descrevendo um objeto astrofísico compacto, cujo campo gravitacional gerado por sua massa impede que mesmo a sua luz emitida escape para o exterior: nasce então o conceito de buraco negro, termo popularizado por John Wheeler nos anos de 1950.\footnote{Cf. o artigo \cite{Saa} sobre os 100 anos da solução de Schwarzschild publicado recentemente na \textit{Revista Brasileira de Ensino de Física}.}

A solução ou métrica\footnote{Solução ou métrica, dentro da relatividade geral, são sinônimos.} de Schwarzschild tem massa, simetria esférica e não possui carga elétrica. É também uma boa aproximação para descrever objetos astrofísicos sem ou com pouca rotação sobre o seu próprio eixo, como o nosso Sol. Objetos imersos num espaço-tempo vazio, sem conteúdo de matéria. Uma similar solução, mas com carga elétrica, foi proposta, independentemente, por Hans Reissner \cite{Reissner} e Gunnar Nordström \cite{Nordstrom} pouco tempo depois. Esta é conhecida como métrica de  Reissner-Nordström ou buraco negro de Reissner-Nordström. E, assim como o buraco negro de Schwarzschild, não tem um movimento de rotação, apresentando, então, a simetria esférica. Somente em 1963 Roy Kerr \cite{Kerr} propôs uma métrica com rotação ou, de forma equivalente, com simetria axial --- nascia então o primeiro buraco negro com rotação no vácuo, a primeira solução das equações de Einstein com tal característica.

Seja no buraco negro de Schwarzschild, ou no de Reissner-Nordström, ou no de Kerr, temos um problema \enquote{aparentemente} sem solução. E um problema nada pequeno. As soluções citadas apresentam uma limitação à teoria de Einstein ou, pelo menos, suas próprias limitações. Tais buracos negros apresentam aquilo que ficou conhecido como uma singularidade. Nesse contexto, uma singularidade significa uma falha, uma \enquote{fissura} nas equações e soluções da relatividade geral. No interior desses buracos negros, a singularidade significa o não funcionamento das soluções. Por exemplo, no centro do buraco negro de Schwarzschild, quando a coordenada radial é $r=0$, a métrica ou solução que o descreve diverge, e quantidades físicas e matemáticas tornam-se incalculáveis, assumem um valor \enquote{infinito}, ou seja, tendem ao infinito. Em Reissner-Nordström, por ter carga elétrica, pode-se descrever o conteúdo de matéria/energia dessa solução (conteúdo dado por um campo eletromagnético que permeia o espaço-tempo) com o uso de um tensor, o chamado tensor energia-momento. E algumas componentes desse tensor, quando $r=0$, divergem. Isto é, dependem da coordenada radial na forma $\sim 1/r^2$.   

Foi somente na década de sessenta do século passado quando uma possível solução começou a surgir para o problema das singularidades no contexto da teoria relatividade geral.\footnote{Digo no contexto da relatividade geral porque há uma suspeita de que uma teoria quântica da gravidade, uma teoria do campo gravitacional quantizado, poderia resolver o problema das singularidades em gravitação. Mas uma teoria completa, confiável e bem aceita pelos físicos em geral ainda não foi apresentada. Mas, como veremos, mesmo no contexto einsteiniano, o da relatividade geral, é possível resolver tal problema.} O russo Andrei Sakharov, num trabalho sobre a formação de estruturas num universo jovem \cite{Sakharov}, obteve um resultado interessante: conforme a matéria se aglomera, devido à gravitação, a densidade de energia não diverge no interior desse aglomerado de matéria, que pode ser uma galáxia em formação. Como veremos na seção III, a não divergência ou não ocorrência de uma singularidade somente é satisfeita caso o espaço-tempo, no interior desse aglomerado, seja um espaço-tempo conhecido como de Sitter, em homenagem ao seu criador,  o holandês Willem de Sitter. Pouco tempo após o resultado de Sakharov, o inglês James Bardeen \cite{Bardeen} utiliza-o e constrói a primeira métrica de buraco negro sem singularidade. A solução ou métrica de Bardeen tem simetria esférica, não possui carga e difere da solução de Schwarzschild por possuir uma massa que não é uma constante $m$ mas uma função $m(r)$, que depende da coordenada radial. Sendo assim, a massa, que na solução de Schwarzschild é pontual e localizada no centro do buraco, em Bardeen espalha-se por todo espaço-tempo. Mas a função $m(r)$ não pode ser uma qualquer. Deve necessariamente fazer com que a métrica de Bardeen seja, no seu núcleo, um espaço-tempo do tipo de Sitter. Com essa exigência, o espaço-tempo no interior do buraco negro é regularizado, sendo então chamado de \textit{buraco negro regular}. Isto é, um objeto compacto, com um horizonte de eventos e sem uma singularidade. O buraco negro de Bardeen foi o primeiro exemplo de um buraco negro sem uma singularidade.\footnote{Cf. \cite{Ansoldi} para uma revisão mais profunda sobre o tema de buracos negros regulares.} E, como veremos, há outros tantos. Porque matematicamente a definição de um buraco negro não envolve a noção de singularidade.  

Neste artigo, apresenta-se uma definição matemática de buraco negro na secção II, com um olhar para as singularidades ou para a sua ausência em tal definição. Em seguida, a seção III trata da métrica de Bardeen e outras soluções de buracos negros regulares. Os comentários finais são apresentados na seção IV. Adotaremos, ao longo deste trabalho, as unidades geométricas: $G=c=1$, sendo $G$ a constante gravitacional universal, e $c$ é a velocidade da luz no vácuo.        

\section{Matemática dos buracos negros}
Matematicamente, buracos negros podem ser definidos com a utilização de conjuntos.\footnote{Usaremos aqui o caminho indicado por um dos textos mais influentes em relatividade geral, o livro de Robert Wald \cite{Wald}.} Para isso, é necessário o conhecimento da chamada estrutura causal do espaço-tempo. Em geometria diferencial --- a área da matemática responsável pelo surgimento da teoria da relatividade geral ---, o espaço-tempo é definido como uma variedade (generalização de superfície) equipada com uma métrica lorentziana (as métricas de Schwarzschild, Bardeen e outras tantas na relatividade geral são métricas lorentzianas porque possuem um determinado número de elementos positivos e negativos em sua diagonal principal, quando são escritas na forma matricial). A métrica --- indicada pelo tensor simétrico $g_{\mu\nu}$ --- dá a medida, o comprimento de vetores e fornece-nos a descrição matemática de um espaço-tempo. Para o cálculo, por exemplo, de distâncias num espaço-tempo qualquer (distâncias infinitesimais entre dois eventos), usa-se o elemento de linha, $ds^2$, que está relacionado à métrica por
\begin{equation}
ds^2=g_{\mu\nu}dx^\mu dx^\nu=g_{tt}dt^2+g_{rr}dr^2+g_{\theta\theta}d\theta^2+g_{\phi\phi}d\phi^2, 
\label{Metrica}
\end{equation} 
com $dx^\mu$ e $dx^\nu$ fazendo o papel de infinitésimos de uma coordenada qualquer (neste trabalho, usaremos as coordenadas esféricas, $x^\mu = \lbrace t,r,\theta,\phi \rbrace$, sendo $t$ a coordenada temporal, e métricas somente com simetria esférica, ou seja, tais quando escritas como uma matriz apresentam somente os seus termos diagonais não nulos). Na relatividade geral, as trajetórias dos corpos são classificadas em três tipos: do tipo tempo, do tipo espaço e do tipo luz. E quando somente a interação gravitacional é levada em conta, tais trajetórias são chamadas de geodésicas. Um corpo que viaja a uma velocidade abaixo da velocidade da luz (sendo esse corpo até mesmo um observador) tem trajetória do tipo tempo. Aquele que viaja mais rápido do que a luz tem a trajetória do tipo espaço. Por fim, a luz percorre uma trajetória do tipo luz, também chamada de trajetória do tipo nula. No que se refere ao elemento de linha, i.e., à distância infinitesimal entre dois eventos num desses três tipos de curvas, para uma trajetória do tipo tempo $ds^2<0$, para uma do tipo espaço $ds^2>0$, e para uma trajetória do tipo luz $ds^2=0$ em nossa convenção.\footnote{Pode-se inverter o sinal de $ds^2$ para curvas do tipo tempo e espaço alterando a assinatura da métrica, que é dada pela quantidade de elementos positivos e negativos em sua diagonal principal.} Outra forma de definir tais curvas ou trajetórias utiliza os seus vetores tangentes. Um vetor qualquer $v$ tem o quadrado de sua norma definido pela métrica na relatividade geral:
\begin{equation}
v^2=g_{\mu\nu}v^\mu v^\nu,
\end{equation}
sendo $v^\mu$ suas componentes. Na trajetória do tipo tempo, o seu vetor tangente tem norma ao quadrado negativa; na do tipo espaço, o seu vetor tangente tem norma ao quadrado positiva; por fim, o vetor tangente a uma trajetória do tipo luz tem norma ao quadrado nula.  

É importante ter em mente os três tipos de trajetórias acima citados para compreender a estrutura do espaço-tempo. Pois, num espaço-tempo simples como o espaço-tempo plano (também conhecido como espaço-tempo de Minkowski\footnote{Vale a pena mostrar a simplicidade do elemento de linha de Minkowski: $ds^2=-dt^2+dr^2+r^2(d\theta^2+\sin^2\theta d\phi^2)$.}), os três tipos de trajetórias têm uma origem e um destino definidos. As do tipo tempo originam-se no infinito passado do tipo tempo ($i^-$) e destinam-se ao infinito futuro do tipo tempo ($i^+$). Da mesma forma, as curvas ou trajetórias do tipo luz --- seus infinitos passado e futuro do tipo luz são $\mathscr{I}^-$ e $\mathscr{I}^+$, respectivamente. Já as trajetórias do tipo espaço têm o infinito do tipo espaço $i^0$. Sendo assim, num espaço plano, a origem e o destino dos corpos (com as suas respectivas trajetórias) estão determinados.\footnote{Nesse ponto, a visão einsteiniana assemelha-se à aristotélica. Aristóteles em \textit{Do Céu} considera que cada corpo tem o seu lugar natural, seja ele fogo, ar, água ou terra. Sendo que a trajetória ou o movimento dos 4 elementos é dirigida aos seus lugares naturais na ausência de forças externas. O elemento terra, abaixo, onde fica o planeta Terra; o fogo fica acima (ou logo abaixo do mundo sub-lunar), e a água e o ar ocupam o espaço intermediário entre a Terra e o mundo sub-lunar.} Mas quando um buraco negro está presente, como veremos, muda-se essa estrutura de infinitos ou a estrutura causal do espaço-tempo. E para a visualização da estrutura causal de espaços-tempo quaisquer, foram desenvolvidos os diagramas de Carter-Penrose. De forma resumida e sem complicações, os diagramas de Carter-Penrose \enquote{trazem} o infinito para o finito. Pois numa finita folha de papel são desenhados os infintos como retas e pontos. Na Fig. 1 são mostrados os diagramas do espaço-tempo de Minkowski e de um espaço-tempo que apresenta um buraco negro em formação. Os infinitos do tipo luz são retas, já os infinitos do tipo tempo e espaço são pontos.\footnote{Para uma introdução e maior compreensão sobre os diagramas de Carter-Penrose, cito o artigo \cite{Coimbra} e o já muito utilizado livro de Sean Carroll \cite{Carroll}, que apresenta o tema no capítulo 5.} 

Com os já conhecidos tipos de infinito, podemos definir um buraco negro, que será indicado por $\mathcal{B}$, para espaços-tempo que são assintoticamente planos, i.e., no \enquote{infinito} esses espaços-tempo são descritos como o espaço-tempo de Minkowski. O espaço-tempo todo, que inclui a região interna e externa ao buraco negro, será indicado por $\mathcal{M}$. Tanto $\mathcal{B}$, $\mathcal{M}$ e os infinitos acima descritos podem ser vistos como conjuntos. Em particular, $\mathcal{M}$ é o conjunto de todos os eventos. Um outro conjunto é necessário para a nossa definição: o conjunto $J^-(\mathscr{I}^+)$. Tal conjunto refere-se a todas as curvas que atingem o infinito luz, $\mathscr{I}^+$, ou seja, os elementos desse conjunto têm uma relação causal com esse infinito futuro, podem afetá-lo num futuro, mesmo que seja num tempo futuro infinito. Dessa forma, $J^-(\mathscr{I}^+)$ é chamado de passado causal do infinito futuro do tipo luz. Sendo assim, um buraco negro terá como definição
\begin{equation}
\mathcal{B}=\mathcal{M}-J^-(\mathscr{I}^+).
\label{Definição}
\end{equation}
Como podemos ver na equação (\ref{Definição}), a definição de um buraco negro (ou a sua região correspondente) exclui do espaço-tempo as trajetórias cujos destinos são o infinito do tipo tempo e do tipo luz e esses dois tipos de infinito. As curvas do tipo tempo e luz do conjunto $\mathcal{B}$ não podem influenciar, mesmo que num tempo infinito, $\mathscr{I}^+$ (e se não podem influenciar  $\mathscr{I}^+$, podem menos ainda influenciar $i^+$). Ou seja, \textit{o buraco negro, a região do espaço-tempo que o define, está desconectado causalmente dos infinitos futuros do tipo tempo e luz, não podendo influenciá-los}. E $\mathcal{B}$ é limitado por uma membrana de mão única --- o famoso horizonte de eventos, uma superfície do tipo luz que pode ser definida como
\begin{equation}
H=\dot{J}^-(\mathscr{I}^+) \cap \mathcal{M},
\end{equation}
onde $\dot{J}^-(\mathscr{I}^+)$ é definido como o contorno do conjunto ${J}^-(\mathscr{I}^+)$. Sendo assim, dentro do horizonte de eventos (indicado na Fig. 1 por $H$, uma reta diagonal, e todas as diagonais nos diagramas são superfícies do tipo luz) não há a possibilidade de corpos, sejam em trajetórias do tipo tempo ou luz, alcançarem o infinito. Ou seja, estão confinados no buraco negro, como podemos ver na Fig. 1, onde o observador B é incapaz de enviar sinais de luz para a região externa ao buraco negro. 

\begin{figure}
\begin{centering}
\includegraphics[scale=0.20]{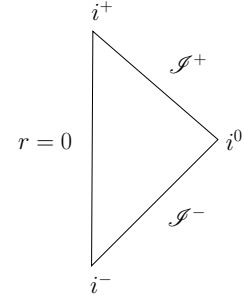} \\
\includegraphics[scale=0.30]{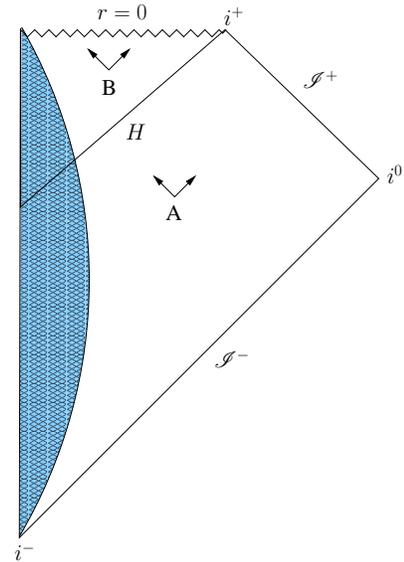}
\par\end{centering}
\caption{Diagramas de Carter-Penrose ou diagramas conforme do espaço-tempo de Minkowski (em cima) e de um espaço-tempo que denota a formação de um buraco negro (embaixo). No espaço-tempo de Minkowski, as curvas têm os seus respectivos destinos  ($i^0,i^+$ e $\mathscr{I}^+$) e origens ($i^0,i^-$ e $\mathscr{I}^-$), indicados pelos três tipos de infinitos. Sendo assim, os diagramas conforme ilustram os três tipos de infinitos usando retas e pontos. Em Minkowski, os três tipos de curvas (do tipo espaço, tempo e luz) seguem os seus caminhos \enquote{naturais}. Mas quando um buraco negro está presente a situação é diferente. Na figura embaixo, o horizonte de eventos de um buraco negro é indicado por $H$. Dentro dele, um observador B emite sinais de luz (que seguem curvas do tipo luz) que não atingem o infinto $\mathscr{I}^+$. Tais sinais dirigem-se diretamente à singularidade, indicada pela linha com formato de serra em $r=0$. Na região externa ao buraco, o observador A pode enviar sinais para dentro do buraco ou para o infinto $\mathscr{I}^+$. Na figura que descreve o buraco negro, a parte escura representa a matéria aglutinando-se para formá-lo. Tal aglutinação é conhecida como colapso gravitacional.}
\end{figure}

No caso do buraco negro de Schwarzschild (e todos aqueles conhecidos na relatividade geral que são métricas ou soluções de um espaço-tempo vazio ou com no máximo um campo eletromagnético) ou mesmo aquele ilustrado na Fig. 1, os corpos que seguem uma trajetória do tipo tempo e do tipo luz, necessariamente, inexoravelmente, dirigem-se à singularidade localizada em $r=0$. Somente corpos que viajam acima da velocidade da luz poderiam escapar ou cruzar o horizonte de eventos para a região externa ao buraco negro.

Para espaços-tempo estacionários (ou seja, que não variam com o tempo) com simetria esférica, a forma do elemento de linha (\ref{Metrica}) apresenta-se explicitamente como
\begin{equation}
ds^2=-f(r)dt^2+\frac{dr^2}{f(r)}+r^2 \left(d\theta^2+\sin^2\theta d\phi^2 \right).
\label{Métrcia_Esférica}
\end{equation} 
No caso específico do buraco negro de Schwarzschild,\footnote{Em Reissner-Nordström, $f(r)=1-\frac{2m}{r}+\frac{Q^2}{r^2}$, sendo $Q$ a carga elétrica do buraco negro.} $g_{tt}=-f(r)=-\left(1-2m/r\right)$, com $m$ fazendo o papel da massa do buraco negro. Sendo assim, fica claro dizer que há uma singularidade nesse espaço-tempo ou nessa métrica. O limite
\begin{equation}
\lim_{r \rightarrow 0}f(r)=\lim_{r \rightarrow 0}\left(1-\frac{2m}{r} \right)=-\infty 
\end{equation}
não é definido. A métrica de Schwarzschild diverge na origem do sistema de coordenadas. E tal divergência não diz respeito ao uso de um sistema de coordenadas particular (em nosso caso, o esférico). Com outro sistema de coordenadas, pode-se notar que a singularidade em $r=0$ permanece. Já a singularidade em $r=2m$, que também faz a equação (\ref{Métrcia_Esférica}) divergir, desparece com a escolha de outro sistema de coordenadas. O raio $r=2m$ em Schwarzschild tem um significado especial. Não denota uma singularidade física, mas o raio do horizonte de eventos. Sendo assim, não diverge, não faz a métrica sofrer dessa \enquote{patologia}. 

Por singularidade, então, pode-se dizer: um ponto, no caso $r=0$ para os exemplos discutidos (Schwarzschild, Reissner-Nordström e o da Fig. 1), que não faz parte de $\mathcal{M}$, o espaço-tempo. A métrica (\ref{Métrcia_Esférica}), como vimos, diverge em $r=0$. Além disso, para esse mesmo ponto, quando se calcula escalares ou grandezas geométricas, tem-se o seu caráter singular reiterado. Por exemplo, o escalar de Kretschmann, $K$, construído a partir do conhecido tensor de Riemann ($R_{\alpha\beta\mu\nu}$), é escrito para a métrica de Schwarzschild como  
\begin{equation}
K=R_{\alpha\beta\mu\nu}R^{\alpha\beta\mu\nu}=\frac{48m}{r^6}.
\label{K_Schwarzschild}
\end{equation} 
Então fica claro, a partir da equação (\ref{K_Schwarzschild}), que o limite de $K$ para $r\rightarrow 0$ não é finito. A chamada singularidade apresenta-se como incomensurabilidade, como uma limitação da descrição dada pela métrica de Schwarzschild. 

Como pôde-se ver, na definição genérica de um buraco negro assintoticamente plano, dada pela equação (\ref{Definição}), não há a menção à singularidade. Somente quando utilizamos o caso particular do buraco negro de Schwarzschild houve uma menção. Mas como veremos, dentro do buraco negro de Bardeen, as trajetórias do tipo tempo e luz dirigem-se ao centro do sistema de coordenadas sem uma singularidade. Temos, então, uma região do espaço-tempo diferente, um pedaço do espaço-tempo semelhante ao espaço-tempo de Sitter. Nascem os buracos negros regulares.              

\section{Buracos negros regulares}
Como foi dito na Introdução, os primeiros passos para a construção de soluções das equações de Einstein sem singularidades foram dados na década de sessenta do século passado. Sakharov \cite{Sakharov}, por exemplo, a partir de um estudo sobre a formação de estruturas num universo jovem em expansão, mostrou que se a densidade de energia da matéria, $\rho$, e a pressão da mesma, $p$, relacionam-se como
\begin{equation}
p=-\rho,
\end{equation}     
que é a equação de estado da métrica de Sitter, a aglomeração da matéria bariônica não produz a divergência de $\rho$. Ou seja, com a aglomeração da matéria devido à gravitação, a densidade de energia não diverge no interior dessa formação. Mas isso somente ocorre no caso em que o espaço-tempo, em seu interior, é do tipo de Sitter. 

O espaço-tempo do tipo de Sitter está entre os mais simples da relatividade geral. Simples em sua forma, pois sua métrica somente difere do espaço-tempo de Minkowski pela adição do termo conhecido como constante cosmológica, $\Lambda$. Esta é a famosa constante que Einstein adicionou às suas equações do campo gravitacional com intuito de obter um universo estático em grandes escalas. Com a observação da expansão cósmica na década de 1920 por Hubble e sua equipe, não sendo o universo mais considerado estático em grandes escalas, Einstein teve que descartar a sua constante. Mas essa teimosa constante retorna na física em 1998 com a observação da expansão acelerada do universo \cite{Supernova,Supernova2}. No modelo cosmológico mais simples, a constante cosmológica é a causa da expansão acelerada do tecido do espaço-tempo, é a origem da chamada energia escura. 

Retornemos à solução de Sitter. Como uma solução da relatividade geral, é simplesmente a solução que descreve um espaço-tempo com simetria esférica, vácuo (sem conteúdo de matéria ordinária, ou escura, ou radiação) e constante cosmológica. Sua forma matemática é
\begin{align}
ds^2 & = -\left(1-\frac{\Lambda}{3}r^2 \right)dt^2+\frac{dr^2}{\left(1-\frac{\Lambda}{3}r^2 \right)} \nonumber \\
& +r^2 \left(d\theta^2+\sin^2\theta d\phi^2 \right).
\end{align}
Como já dissemos, $\Lambda$ é a constante cosmológica, que pode ser positiva ou negativa: quando positiva, o espaço-tempo é de Sitter; quando negativa, é anti-de Sitter. Não apenas na física de buracos negros a solução de Sitter é importante. Em cosmologia, a chamada fase inflacionária, onde o universo teve uma expansão acelerada logo depois do suposto \textit{big bang},\footnote{Suposto pois hoje são possíveis modelos cosmológicos sem a singularidade inicial ou o \textit{big bang}. As cosmologias com ricochete surgem como opções na ciência atual, são alternativas ao problema das singularidades mesmo dentro da teoria da relatividade geral. Para uma introdução, veja \cite{Novello1}. Já para um estudo mais profundo, a revisão \cite{Novello2} é indicada.} é descrita como um período onde o espaço-tempo é \textit{quase} de Sitter ($p\simeq -\rho$). Já o espaço-tempo anti-de Sitter é importante para a, hoje muito estudada, correspondência AdS-CFT (\textit{Conforme Field Theory in anti-de Sitter Spacetime}).

Agora que temos uma ideia do que é um espaço-tempo de Sitter, podemos entender o que foi dito acima sobre os buracos negros regulares. Retornemos à métrica (\ref{Métrcia_Esférica}) com simetria esférica, a que descreve um buraco negro sem rotação. Tal solução descreve tanto a solução de Schwarzschild quanto a de Bardeen: quando a massa do buraco negro é constante, $f(r)=1-2m/r$, temos Schwarzschild; quando $f(r)=1-2m(r)/r$, e a função $m(r)$ tem uma forma determinada, ou seja, é uma função representada pela equação
\begin{equation}
m(r)=\frac{Mr^3}{\left(r^2+e^2\right)^\frac{3}{2}},
\label{Função de Massa}
\end{equation} 
temos o buraco negro regular de Bardeen. Na equação (\ref{Função de Massa}), $M$ e $e$ são constantes: a primeira é interpretada como um parâmetro de massa, e a segunda, como veremos, é tida como um tipo de carga. A adoção de uma função para a massa --- ao invés de considerá-la uma constante --- produz algumas diferenças entre as soluções de Bardeen e Schwarzschild. Não apenas no que diz respeito ao problema da singularidade. Em Schwarzschild, há uma superfície do tipo luz importante como vimos: o horizonte de eventos. Tal superfície, que funciona como uma membrana de mão única, pode apresentar-se em dobro no buraco negro de Bardeen. Dependendo da relação entre $M$ e $e$ há a possibilidade de um horizonte interno e um horizonte externo (sendo o último um horizonte de eventos como no buraco negro de Schwarzschild).  

Ora, para observar o desparecimento da singularidade e a solução desse \enquote{terrível} problema com a adoção da equação (\ref{Função de Massa}), usa-se uma aproximação para a função da massa, para $r$ pequeno, dada por
\begin{equation}
m(r)\approx M \left(\frac{r}{e} \right)^3, 
\end{equation}
que conduz à 
\begin{equation}
f(r)\approx 1-Cr^2,
\end{equation}
sendo $C=2M/e^3$ uma constante positiva. Com essa aproximação obtida para $f(r)$, substituindo-a na equação (\ref{Métrcia_Esférica}), a métrica de Sitter é obtida para valores pequenos de $r$. Ou seja, com o uso da função de massa de Bardeen, a métrica (\ref{Métrcia_Esférica}) que descreve um buraco negro esférico apresenta um núcleo, uma região interna, do tipo de Sitter. Uma região para valores pequenos da coordenada radial $r$ onde o espaço-tempo apresenta-se como de Sitter.  

O simples \enquote{truque} matemático feito por Bardeen (a substituição de $m$ por uma determinada função de massa) tornou a métrica (\ref{Métrcia_Esférica}) regular, removeu a singularidade situada na origem do sistema de coordenadas $r=0$, fazendo deste ponto um ponto qualquer de $\mathcal{M}$. A regularidade da solução de Bardeen fica clara quando se observa diretamente a métrica ou se calcula escalares. Por exemplo, o escalar de Kretschmann para a métrica de Bardeen exemplifica a sua regularidade:
\begin{equation}
\lim_{r\rightarrow 0} K=\lim_{r\rightarrow 0} R_{\alpha\beta\mu\nu}R^{\alpha\beta\mu\nu}= 96\left(\frac{M}{e^3} \right)^2.
\end{equation}
Nesse caso, ao contrário do escalar de Kretschmann da métrica de Schwarzschild, dado pela equação (\ref{K_Schwarzschild}),  o limite para $r$ tendendo a zero é finito. E, igualmente, a métrica também apresenta-se regular, sendo $\lim_{r\rightarrow 0}ds^2$ finito. Sendo assim, o buraco negro de Bardeen mostra-se como regular, sem possuir uma singularidade na origem do sistema de coordenadas. E mostra-se como buraco negro, acima de tudo, por possuir, no mínimo, um horizonte de eventos.  

Décadas após a sua publicação, a métrica de Bardeen foi (e continua sendo) alvo de investigações. Em \cite{Beato} mostra-se que o buraco negro de Bardeen tem uma origem, isto é, pode-se interpretá-lo como uma solução exata das equações do campo gravitacional. E exata, nesse caso, significa uma solução com uma fonte determinada. Na solução de Bardeen, a fonte --- de acordo com Ayon-Beato e Garcia, que interpretaram $e$ como um tipo de carga, i.e., um monopolo magnético --- vem de uma eletrodinâmica não linear. Com uma eletrodinâmica não linear acoplada à relatividade geral, ideia expressa pela ação 
\begin{equation}
\mathcal{S}= \int dv \left(\frac{1}{16\pi}R - \frac{1}{4\pi}\mathcal{L}(F)\right),
\label{Ação}
\end{equation}
sendo $R$ o escalar de Ricci, e $\mathcal{L}(F)$ é uma complicada densidade lagrangiana não linear,\footnote{A expressão para $\mathcal{L}(F)$ é $\frac{3}{2se^{2}}\left(\frac{\sqrt{2e^{2}F}}{1+\sqrt{2e^{2}F}}\right)^{\frac{5}{2}}$, com $F=\frac{1}{4}F_{\mu\nu}F^{\mu\nu}$. $F_{\mu\nu}$ faz o papel do tensor eletromagnético, enquanto $s$ é uma constante dada por $\vert e \vert/2M$.} as equações de Einstein são obtidas com um tensor energia-momento não nulo. Tal tensor descreve uma eletrodinâmica não linear como fonte da solução de Bardeen (quando o segundo termo da equação (\ref{Ação}) é nulo, as equações de Einstein são obtidas no vácuo assim como as suas soluções sem conteúdo de matéria, como a de Minkowski e a de Schwarzschild). 

Mas as pesquisas em buracos negros regulares vão além. Hoje há outras soluções ou métricas regulares disponíveis na literatura. Sean Hayward \cite{Hayward}, por exemplo, construiu uma solução regular, similar à de Bardeen, com outra função $m(r)$ com o intuito de descrever a formação e evaporação de buracos negros regulares. Há também as soluções com simetria axial, as que descrevem buracos negros regulares com rotação. Nosso trabalho \cite{Neves_Saa} tratou desse tema, e buracos negros regulares com rotação foram obtidos, além disso, com a adoção da famosa constante cosmológica e uma função de massa geral, que abrange as funções utilizadas por Bardeen e Hayward. E não apenas no contexto da relatividade geral buracos negros regulares são estudados. Mesmo em teorias que são propostas para substituir a gravitação einsteiniana ou a física de hoje, como a gravidade quântica em \textit{loops} ou os mundos branas,\footnote{Em mundos branas, o nosso trabalho \cite{Neves} discute buracos negros regulares com ou sem rotação.} buracos negros regulares são previstos.

\section{Comentários finais}
Ao contrário do que se pode pensar, um buraco negro não precisa necessariamente conter uma singularidade. Como vimos, na definição matemática de um buraco negro, apenas o horizonte de eventos é mencionado como aquilo que lhe é inerente. Ou seja, para que um objeto astrofísico seja reconhecido como buraco negro, apenas a membrana de mão única, o horizonte de eventos, deve ser levada em conta. Como uma consequência dos trabalhos de Andrei Sakharov e seus colaboradores, o expediente para construir matematicamente os buracos negros sem uma singularidade surge já em 1968 com James Bardeen. Os resultados da Sakharov mostravam a possibilidade de evitar o problema das singularidades mesmo no contexto da relatividade geral por meio de um tipo de espaço-tempo: o espaço-tempo do tipo de Sitter. Com isso, com tal espaço-tempo no interior de um buraco negro, é possível evitar o aparecimento de uma singularidade no interior de um objeto astrofísico, assim como fez Bardeen. Surgem, então, os buracos negros sem uma singularidade ou os buracos negros regulares --- tema atual na física hoje. 

\section{Agradecimentos}
Gostaria de agradecer à FAPESP (Fundação de Amparo à Pesquisa do Estado de São Paulo) pelo apoio financeiro (processo número 2013/03798-3).

\end{document}